\documentclass[manuscript]{aastex}

\usepackage{graphicx}
\usepackage{rotating}
\usepackage{pdflscape}
\usepackage{color}
\usepackage{lineno}
\linenumbers*[1]
\usepackage{mathtools}
\usepackage{amsmath}
\usepackage{float}
\usepackage{color}

\newcommand{\red}{\textcolor{black}}

\newcommand{\beq}{\begin{equation}}
\newcommand{\eeq}{\end{equation}}
\newcommand{\baq}{\begin{eqnarray}}
\newcommand{\eaq}{\end{eqnarray}}

\shorttitle{Spectral hardening of solar flares}
\shortauthors{G. Li, X. Kong, G. Zank, Y. Chen}

\begin{document}
\title{On the spectral hardening at $\sim >$ $300$ keV in solar flares}

\author{G. Li$^{1,*}$, X. Kong$^{2,1}$, G. Zank$^{1}$, Y. Chen$^{2}$ }
\affil{  { $^1$ Department of Physics and CSPAR, University of
Alabama in Huntsville, Huntsville, AL 35899, USA  \\
$^2$
Institute of Space Sciences and School of Space Sciences and Physics, 
Shandong University, Weihai, China, 264209 } \\
$^*$ gang.li@uah.edu }

\begin{abstract}
It has been noted for a long time that the spectra of observed continuum emissions in
many solar flares are consistent with double power laws with a hardening at energies
$\sim > $ 300 keV.  It is now largely believed that at least in electron-dominated events
the hardening in photon spectrum reflects an intrinsic hardening in the source electron spectrum.
In this paper, we point out that a power law spectrum of electron with a hardening at high
energies can be explained by diffusive shock acceleration of electrons at a
termination shock with a finite width.
Our suggestion is based on an early analytical work
by Drury et al., where the steady state transport equation at a shock with a tanh profile was solved
for a $p$-independent diffusion coefficient. Numerical simulations with a $p$-dependent diffusion
coefficient show hardenings in the accelerated electron spectrum which are comparable
with observations. One necessary condition for our proposed scenario to work is that
high energy electrons resonate with the inertial range of the MHD turbulence and low energy electrons
resonate with the dissipation range of the MHD turbulence at the acceleration site, and the spectrum of 
the dissipation range $\sim k^{-2.7}$. A $\sim k^{-2.7}$ dissipation range spectrum is
consistent with recent solar wind observations.
\end{abstract}


\maketitle

\section{Introduction}

Our Sun is an efficient particle accelerator. Ions with energy up to $\sim GeV$/nucleon are detected 
in-situ during large Solar Energetic Particle (SEP) events where both large flares and fast CMEs often
occur together. At flares, electron Bremsstrahlung is believed to be the main source of
the continuum radiation  (e. g. \citet{Ramaty.etal75, Vestrand88}).

Continuum emissions provide invaluable information that constrains the underlying acceleration mechanism.
These constraints include the energy budget, the total number of electrons, the acceleration time
scales, etc. See reviews by \citet{Miller.etal97}  and \citet{Zharkova.etal11} for a detail discussion on
various acceleration mechanisms and the implications of these constaints on them.

One observational constraint that received less attention, even though has been noted for a long time,
 is the hardening of the continuum spectrum at high energies (often $\sim > 300$ keV). 
In 1975, \citet{Suri.etal75} examined the X-ray and gamma-ray flux in the August 4, 1972 event and concluded 
that the X and gamma-ray flux was produced by a single population of electrons with a break
in its spectrum, instead of two separate populations acting independently.
Later,  {\citet{Yoshimori.etal85} }, using Hinotori spacecraft, examined the hard X-ray (HXR) 
spectrum in a broad energy range (20 keV - 7 MeV) for four flares that showed significant
hardening at energies above $\sim $ 400 keV. They confirmed the earlier suggestion of  \citet{Suri.etal75} 
that the hardening in the continuum reflects an underlying hardening in the source electron spectrum. 
Spectral hardening also occur in events where there are clear signatures of gamma-ray lines.
 The most recent report of such an event is from the Fermi observation (see \citet{Ackermann.etal12}) where
spectral hardening was found at {above several hundred keV.} 

Note that hardening in the source electron spectrum is not the only cause for a hardening in the photon spectrum. 
Various processes, such as electron-electron Bremsstrahlung, proton Bremsstrahlung, positronium annihilation 
continuum and inverse Compton emissions (see \citet{Vestrand88}), may all
lead to some hardening of a continuum emission from a straight power law (without hardening) source electron 
spectrum.  However, for parameters appropriate to a solar flare site, the contributions of these
 processes are relatively small and the resulting hardening of the spectral index is perhaps
$\sim 0.5$  { \citep{Kontar.etal07}}. Therefore these processes can not explain events where 
the change of spectral indices are as high as $2$. \citet{Park.etal97} studied
photon spectral hardening around 1 MeV for four flares. In their scenario, the emission is a simple sum of the 
thin target emission from the trapped electrons at the acceleration site near the loop
top and the thick target emission from the escaping electrons precipitating on the solar surface. 
With the assumption that electrons having smaller energies have shorter escape times, and
noted that the energy dependence of the Bremsstrahlung cross section differs in the nonrelativistic and the 
relativistic regimes, \citet{Park.etal97} were able to account for the observed spectral
hardening. Note in the scenario of \citet{Park.etal97}, the energy dependence of the escaping time decides 
the hardening, and the accelerated electron spectrum in the accelerated region does not
(need to) have a spectral hardening. In-situ observations by
\citet{Moses.etal89}, however, showed that electron spectral
hardening is rather common in short duration events.  Assuming
these in-situ electrons are the source electrons escaping from the acceleration site through
interchange reconnection, then \citet{Moses.etal89}'s results also suggest that 
the accelerated electron population has a hardening at high energies.

\red{
High energy electrons also lead to microwave emissions through gyro-synchrotron radiation (see e.g. the 
recent review by  \citet{White.etal11} for a detailed discussion of the relationship between solar radio 
and HXR emissions).
In an early study, using BATSE (HXRs) and Owens Valley Radio Observatory (microwaves), \citet{Silva.etal00}
examined 27 solar flares with multiple peaks (a total of 57) which were observed at both HXR and
microwave wavelengths. Fitting the HXR spectra by a single power law and the microwave spectra as 
gyrosynchrotron emissions, \citet{Silva.etal00} found that in 75\% of the bursts, the 
inferred spectral indices of the electron energy distribution of the microwave-emitting
electrons were harder (by $0.5$ to $2.0$) than those of the lower energy HXR emitting electrons. 
\citet{Silva.etal00} concluded that there exists a breakup in the energy spectra of the source electrons at 
around $\sim 300$ keV,  in agreement with previous observations of HXR-alone spectra of giant flares.  }

\red{
Note, however, in most events the HXRs are emanated from the footpoints of flare loops
 and microwaves are emanated from the tops of flare loops. Furthermore, there is also a delay between the 
peak of the microwave emission and the HXR emission. So there are transport effects between
electrons generating microwave emissions and those generating HXRs \citet{White.etal11}.
Both the harder HXR spectrum at the footpoints and the delays of the microwave emission could be caused by 
magnetic trapping of higher energy electrons near the 
looptop and the precipitation of lower energy electron to the footpoints, as first suggested 
by \citet{Melrose.Brown76}. In a recent study by \citet{Kawate.etal12}, HXR and microwave emissions 
from 10 flares were analyzed. 
Although the emissions were at different locations and the spectral indices for microwave emissions are 
harder than those of the HXRs, by assuming a spectrum for the accelerated electrons that 
is consistent with the HXR emissions (but extend to higher energies), \citet{Kawate.etal12} were able to 
produce microwave spectra comparable to the observations. The authors concluded that it is a single electron 
population that is responsible for the HXRs and microwaves emissions and the hardening of the microwave emission 
is  due to a more efficient trapping of electrons with higher energies.
In another study,  to minimize the effect of the trapping of high energy electrons on the resulting 
spectra of looptop microwave emissions, \citet{Asai.etal13} examined both the HXR and microwave spectra 
prior to the peak emission in 12 flares. They still find a significant hardening of the source electrons for 
the microwave emissions. These authors suggest that there is an intrinsic spectral hardening for the 
source electron spectrum around several hundreds of keV and the microwave gyrosynchrotron emission is 
due to electrons at higher energies (in the harder part of the spectrum). 
}

\red{ In this work, we do not consider microwave emissions and focus on HXRs alone.}
\citet{Vestrand.etal99} identified $258$ flare events using SMM observations. 
Among these, many are  electron-dominated events with no clear signature of gamma-ray emissions
\citep{Rieger.Marschhauser90,Marschhauser.etal94}. In these
events, the contribution of nuclear gamma ray lines is minimal and the
 continuum is mainly due to Bremsstrahlung of the energetic electrons. The spectral hardening can be clearly 
seen in many of those electron-dominate events. 
A careful examination of these events based on the mechanism proposed here will be reported 
elsewhere (Kong et al. to be submitted).

A hardening in the source electron spectra is hard to explain for any acceleration mechanism.
In this paper, we propose a scenario which is based on diffusive shock acceleration (DSA) to explain
the observed spectral hardening.

Electron acceleration at a termination shock (TS) in solar flares is not a new idea.
\citet{Tsuneta.Naito98} were the first to consider electron acceleration via DSA 
at a flare TS. \citet{Tsuneta.Naito98} pointed out that slow shocks bounding the reconnection 
X-point can heat the plasma up to perhaps $10$-$20$ MK, providing abundant seed population which 
are accelerated to 1 MeV in $0.3$ to $0.6$ seconds at the TS.

Noting that the standing TS is of quasi-perpendicular in nature,
 \citet{Mann.etal09} considered shock drift acceleration (SDA) at a standing TS.
In the work of \citet{Mann.etal09}, most energy gain of electrons is through a single reflection at the shock
front. Therefore to accelerate electrons to high energies, a stringent requirement of $\theta_{BN}$
(e.g. $\theta_{BN} > 88^\circ$) is needed.
However, while the TS on large scale is of quasi-perpendicular, small scale structures (such as ripples)
exist on the shock front. Indeed, the plasma in the reconnection region
is unlikely to be homogeneous, so the resulting TS is unlikely to be planar.
Recently, \citet{Guo.Giacalone12} have examined electron acceleration at a flare TS using a
hybrid code. In the simulation of \citet{Guo.Giacalone12}, 
 many small scale ripples were identified along the shock surface.
The existence of these ripples suggests that assuming a shock with 
a $\theta_{BN} > 88^\circ$ across the shock surface may be unrealistic. 
Furthermore, the existence of these small scale structures implies that one single field line can intersect 
the shock surface multiple times. Consequently, the acceleration process will be of diffusive in nature.
In this work, we follow  \citet{Tsuneta.Naito98} and assume the electron acceleration at
a flare shock can be described by the DSA mechanism.
Note that the existence of a TS in a flare site is not trivial.
Observational evidence of flare TS has been reported by \citet{Warmuth.etal09}, 
who used dynamic radio spectrum from the
Tremsdorf radiospectrograph to show that there was a type-II radio bursts from a standing TS at
$\sim 300$ MHz during the impulsive phase of the X1.7 flare of 2001 March 29.
\red{ Besides the existence of a TS, the area of the TS shock has also to be large enough
 ($\sim 10^{20}$ cm$^2$ in large flares ) to account for the observed flux of HXRs generated by high 
energy electrons. By assuming a 50\% contour of the NRH source at 327 MHz (see Figure 2 of \citep{Warmuth.etal09} ) being 
a proxy for the shock, \citet{Warmuth.etal09} estimated a shock area  of 
$ \sim 1.3$*$10^{20}$ cm$^2$ in the 2001 March 29 X1.7 flare.  
We do note that these areas are much larger than the areas of HXR sources and are more comparable
to active region sizes. 
Whether or not the size of the TS can be this large remains to be examined. 
Using the same technique, \citet{Warmuth.etal09} nevertheless obtained similar areas for other events where 
TS were observed.
Of course, for smaller flares (like M flares), the area of the active region is smaller and we 
expect the area of the shock is also smaller. }

Besides shock acceleration, models based on stochastic (aka 2nd-order Fermi) acceleration exist.
For example, \citet{Miller.etal96,  Miller.etal97}  assumed the presence of some large scale turbulence at
the flare site and considered the coupled system of the wave cascading and particle acceleration.
\citet{Miller.etal96,  Miller.etal97} showed that various modes of waves (Alfv\'{e}nic and fast mode waves),
as they cascade to small scales,  can efficiently accelerate both ions and electrons.  Similar processes have
also been studied by, for example, \citet{Petrosian.etal94, Park.etal97}. Unlike \citet{Miller.etal96,  Miller.etal97},
\citet{Petrosian.etal94, Park.etal97} did not address the cascading of the turbulence and assumed the wave
spectra is given. 

In this work, we do not consider stochastic acceleration. However, as in \citet{Petrosian.etal94, Miller.etal96,
Miller.etal97, Park.etal97}, we assume the 
diffusion coefficient $\kappa$ is decided by the underlying turbulence power at a flare site.

\section{Diffusive shock acceleration of electrons at a finite-width termination shock}
At a piecewise shock, the standard steady state DSA predicts a power law spectrum $\sim p^{-\alpha}$ for
energetic particles. The power law spectral index $\alpha$ is given by $3s/(s-1)$, where
{ $s = u_{1}/u_{2}$ is the compression ratio,  $u_{1}$ and $u_{2}$ the upstream and downstream flow speed in 
the shock frame}. In the case
of a shock having a finite width $\sim L_{diff}$, \citet{Drury.etal82} showed that the spectral index depends 
on the shock width. Assuming the background fluid speed is given by a tanh profile: 
\beq 
u(x) = \frac{u_1+u_2}{2}
-\frac{u_1-u_2}{2} tanh(x/L_{diff}) 
\label{eq:u} 
\eeq
then the spectral index $\alpha$ becomes \citep{Drury.etal82}, 
\beq 
\alpha = \frac{3s}{s-1} (1+ \frac{1}{\beta_s} \frac{1}{s-1})
\label{eq:gamma} 
\eeq 
where $\beta_s$ is a dimensionless parameter
and is related to the diffusion coefficient $\kappa$ through, 
\beq
\kappa = \beta_s (u-u_1) (u-u_2) \frac{dx}{du} = \beta_s
\frac{u_1-u_2}{2} L_{diff}. \label{eq:kappa0} 
\eeq
Although \citet{Drury.etal82} considered only the case of $p$-independent $\kappa$ where
analytical solutions can be obtained, one can see from the above that
for a $\kappa$ increasing with $p$ the spectrum will harden at high energies.
Because the factor of $1/\beta_s$ in equation~(\ref{eq:gamma}), the spectral index
quickly approaches the limit of $3s/(s-1)$ when $\beta_s \ge 1$.
When $\beta_s$ is small, however, the second term in the bracket of equation~(\ref{eq:gamma}) 
dominates and the spectrum can be very soft.

Clearly the momentum dependence of the diffusion coefficient $\kappa$ decides the shape of the spectrum.
At a flare site, the $\kappa$ of energetic electrons is decided by the turbulence level.
At large scales, the turbulence is of Alfv\'{e}nic and particle-wave cyclotron resonance
can accelerate ions to high energies via the stochastic acceleration process (e.g. \citet{Miller.etal97}).
 For electrons, except at very high energies, however, they
do not resonate with Alfv\'{e}n waves, therefore they interact with other waves, for example, 
fast mode and/or whistler waves \citep{Miller.etal96}.

\red{Note that  \citet{Drury.etal82} did not consider the effect of the energetic electrons on the shock.  
In a more refined and self-consistent analysis, the pressure of the energetic electrons 
needs to be taken into account and it will affect the shock width. 
 This is similar to a modified shock structure caused by energetic cosmic rays as first examined by 
\citet{Axford.etal82}.  Such a discussion, however, exceeds the scope of this work and we do not 
consider the back reaction of energetic electrons on the shock structure.}

We assume the turbulence at a flare, as in the solar wind, is described by an inertial range
joining to a dissipation range and the power density $I(k)$ is given by,
\beq 
I(k) =  I(k_0) ( (\frac{k}{k_b})^{-\epsilon_i} H(k_b -k)  +
(\frac{k}{k_b})^{-\epsilon_d} H(k-k_b) ),
 \label{Ik} \eeq 
where $\epsilon_i$ and $\epsilon_d$ are the spectral indices in the inertial range and dissipation range, 
respectively. We assume $\epsilon_d = 2.7$ (see below) and consider three cases for
$\epsilon_i$: $5/3$, $1.5$ and $1.0$. The case of $\epsilon_i = 5/3$ corresponds to a Kolmogorov cascading; 
the case of {$\epsilon_i = 1.5$} corresponds to a Iroshnikov-Kraichnan (IK) cascading,
and the case of  {$\epsilon_i = 1.0$} corresponds to a Bohm-like diffusion (see below). 
At very small $k$, the energy containing range sets in and $I(k)$ bent over. The normalization
of $I(k)$ is given by, \beq \int_{-\infty}^{+ \infty} I(k) =
<\delta B^2>. \eeq For a wide range of electron energy, the
resonating wavenumber $k$ is in the dissipation range. 
In the solar wind, one finds a spectrum $\sim k^{-2.7}$ to $\sim k^{-3.0}$ in the dissipation range
\citep{Leamon.etal98, Leamon.etal99, Chen.etal10, Howes.etal11, Alexandrova.etal12}.
Unlike the inertial range, the nature of the turbulence in the dissipation range is still under
debate. Two possible scenarios include Landau damping of kinetic
Alfv\'{e}n waves  { e.g. \citep{Leamon.etal99, Leamon.etal00, Boldyrev.Perez12} }, or
 whistler waves  { e.g. \citep{Stawicki.etal01, Krishan.Mahajan04, Galtier06} }.
For KAWs, $k_{\perp} \gg k_{||} $, electron-wave interaction is
through the Landau resonance and KAWs can effectively heat
electrons. Whistler waves have $ \omega_{p} < \omega  < \Omega$ and electrons can interact with 
whistler waves through the cyclotron resonance.  The resonance condition is, 
\beq \omega
- k_{||} v_{||} = n \Omega 
\label{eq:resonance}
\eeq 
where $\Omega = eB/ (\gamma m_e)$ is the electron cyclotron frequency and $\gamma$ is the
Lorentz factor. For low frequency waves $\omega < \Omega$, the
resonance condition (on taking $n=1$) yields \beq
 \mu v =  \Omega / |k_{||}|
\eeq
where $\mu$ is the pitch angle of the electron.  Note from equation~(\ref{eq:resonance}), one can 
see that when the energy of electron is high enough, it can also resonate with Alfv\'{e}n waves.
In this work, we assume the dissipation range turbulence is whistler-wave-like and 
electrons can resonate with the wave through the cyclotron resonance. 
As done in \citet{Gordon.etal99, Rice.etal03, Li.etal05}, we further simplify the resonance condition by
replacing $k=\Omega/\mu v$ with  $k=\Omega/ v$, which corresponds to an extreme resonance broadening.
The pitch angle diffusion coefficient $D_{\mu \mu}$, from the Quasi-linear Theory (QLT) \citep{Jokipii66}
is,
\beq
D_{\mu \mu} = \frac{1- \mu^2}{|\mu| v} \frac{\Omega^2}{B_0^2} I(k=\Omega/\mu v)
\label{Duu}
\eeq
The diffusion coefficent $\kappa$ is related to $D_{\mu \mu}$ through,
\beq
\kappa = \frac{v^2}{8} \int_{-1}^{+1} \frac{(1-\mu^2)^2}{D_{\mu \mu}} = \frac{v^3 B_0^2}{16 \Omega^2 I(k=\Omega/v)}
\label{kappa1}
\eeq

We make no attempts to estimate the turbulence level at the reconnection site in this work. Instead, we are more
interested in the energy dependence of $\kappa$. From equation~(\ref{kappa1}), we have
\beq
\kappa = \kappa_0 \frac{(p/p_0)^{3-\epsilon_{i,d}}}{\gamma}
\label{kappa2}
\eeq
where subscripts $i$ or $d$ denote whether electrons resonate with the inertial or the dissipation range 
of the turbulence. For electrons resonating with the dissipation range that have
a $\epsilon_d \sim 2.7$, equation~(\ref{kappa2}) suggests that
$\kappa$ has a very shallow dependence on electron momentum (energy).  In comparison,
for electrons resonating with the inertial range, $\kappa$ increases quickly
with particle momentum (energy). In \citet{Tsuneta.Naito98}, the Bohm diffusion approximation was used,
in which case $\kappa \sim v R_l$, where $R_l$ is electron's gyroradius. 
This corresponds to an $\epsilon_i = 1$.

\begin{figure}[ht]
\includegraphics[width=0.8\textwidth]{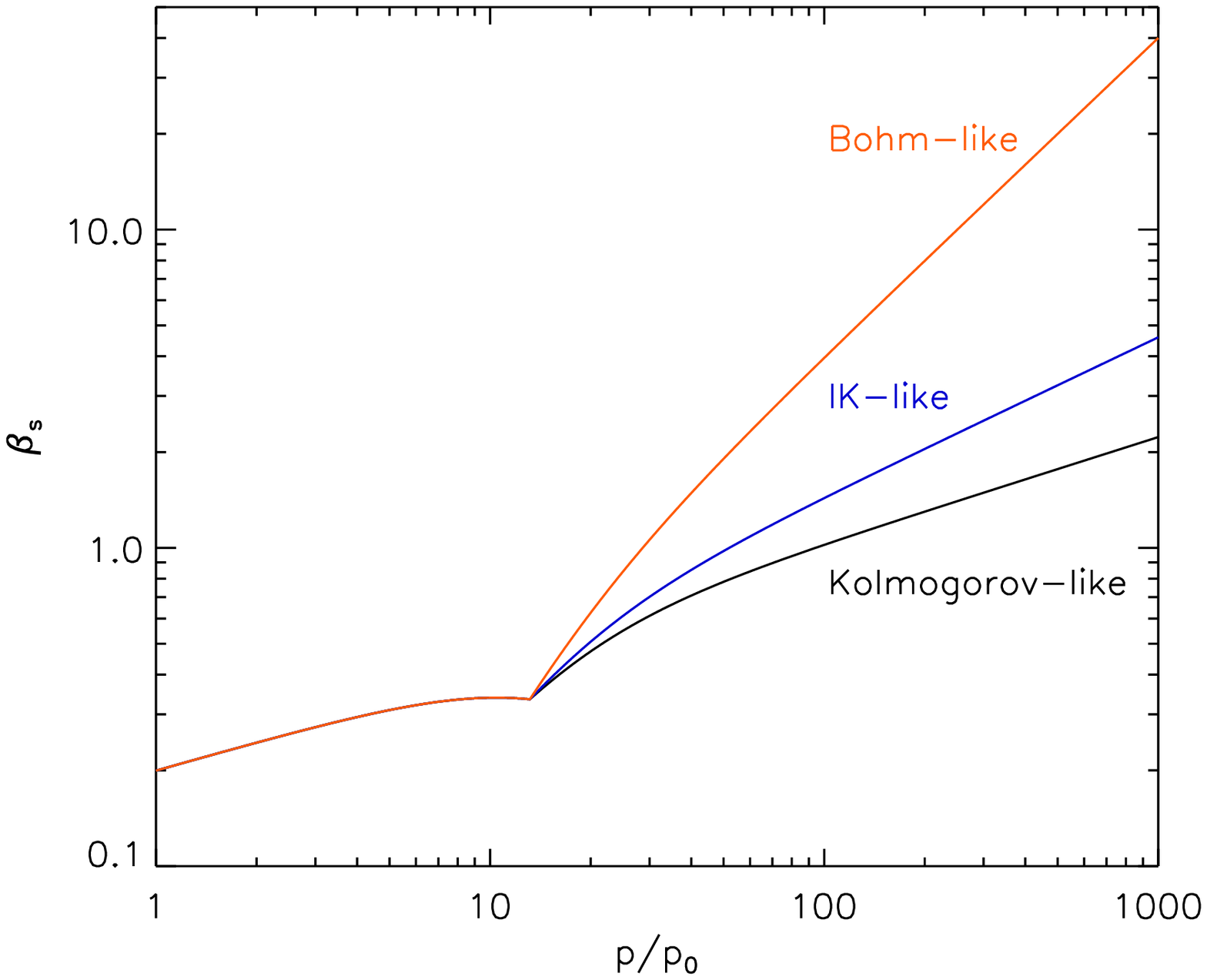}
\caption{ $\beta_s$ as defined in equation~(\ref{eq:kappa0}). At low energies, electrons resonate with the dissipation
range and $\beta_s$ has a weak momentum dependence. Above $p_b$, electrons resonate with the inertial range and 
$\beta_s$ quickly increases with momentum. Three cases for the inertial range are considered. These are, from top to 
bottom, Bohm-like, IK-like and Kolmogorov-like.  }
\label{fig-beta}
\end{figure}

The fact that $\kappa$ has a very shallow dependence on the electron's momentum in the dissipation range 
and a strong dependence in the inertial range is the key to understand the hardening of electron spectrum.
In Figure~\ref{fig-beta} we plot $\beta_s$ as defined in equation~(\ref{eq:kappa0}), where 
from equation~(\ref{kappa2}), we have
\beq
\beta_s = \beta_0 \frac{(p/p_0)^{3-\epsilon_{i,d}}}{\gamma}
\label{beta}
\eeq
We set $\beta_0=0.2$. This value yields an electron spectral index at low energy to be $\sim p^{-10}$, comparable
to flare observations.

Note, from equation~(\ref{eq:kappa0}),  $\beta_s$ also depends on the width of the shock.
Simulations by \citet{Scholer.Burgess06} suggested that the shock width is of the order of
ion inertial scale length $\sim (c/\omega_{pi})$. On the other hand, observations of the Earth's bow shock
(at quasi-perpendicular configurations) showed that its ramp width is somewhat smaller than $\sim (c/\omega_{pi})$
\citep{Scudder.etal86, Balikhin.etal95, Newbury.etal98}.
In particular, \citet{Newbury.etal98} found considerable fine structures of the order of
 $\sim (c/\omega_{pe})$.  {\citet{Zank.etal01}} suggested that these fine structures will help to circumvent the
injection problem for Anomalous cosmic rays. In a very
recent study, using Clusters observation, \citet{Schwartz.etal11} showed that at the Earth's bow shock half of
the temperature occurred in about $\sim 7 c/\omega_{pe}$ or $\sim (1/7) c/\omega_{pi}$. The total width of the shock in
 {\citet{Schwartz.etal11}}, which is close to $L_{diff}$ in our work, however, is another factor of 
$\sim 6$ (see their figure (3)).  Therefore, in this work, we assume the shock width is given by the 
ion inertial length scale $L_{diff} \sim c/\omega_{pi}$.

The break point $p_b$ in Figure~\ref{fig-beta} is $p_b \sim \gamma m_e \Omega/k_b $ with $k_b$ the wave number
separating the inertial range and the dissipation range. The scale at which 
the inertial range transits into the dissipation range is still a much debated issue. 
It has been argued that it could be  the thermal proton Larmor radius $\sim \frac{\sqrt{k_B
T/m_p}}{\Omega_p}$  { \citep{Leamon.etal98, Leamon.etal99} } or
the ion inertial length  $\sim \frac{V_A}{\Omega_p}$  with $\Omega_p$ the proton cyclotron frequency
{\citep{Leamon.etal00, Smith.etal01}}. 
Consider a typical flare site \citep{Miller.etal96, Mann.etal09} with a temperature
of $ T \sim 5$ MK, a magnetic field of $B \sim 200$ Gauss and a density of $n_{e} \sim 10^{9}$ cm$^{-3}$,
we find 
an Alfv\'{e}n speed $V_A \sim 1.38* 10^4$ km sec$^{-1}$,  
a thermal proton speed $v_{th} \sim 200$ km sec$^{-1}$, 
a proton gyrofrequency $\Omega_p = 1.91 * 10^6$ Hz.  
Consequently, the thermal ion Larmor radius is $\sim 0.10$ m and  the ion inertial length is  $\sim 7.2$ m.
If $k_b$ is the reciprocal of the thermal ion Larmor radius, then $p_b \sim 0.64$ MeV/c and the corresponding
kinetic energy is $0.31$ MeV. This is in good agreement to the observed continuum
 emission break 
locations $\sim 300$ keV. On the other hand, if $k_b$ is the reciprocal of the ion inertial length, 
then $p_b \sim 39 $ MeV/c, much too high for the proposed scenario. Therefore our proposed scenario favors the 
the suggestion of \citet{Leamon.etal98, Leamon.etal99} that the dissipation range sets in at the thermal 
ion Larmor radius scale. 
Comparing to the width of the shock, which is the ion inertial length scale $7.2$ m,
the gyro-radius of an electron
  { $R_l = \gamma v / \Omega_e$ }is $0.1$ ($0.4$) m for a kinetic energy of $300$ keV ($2$ MeV).

Using a momentum dependent $\beta_s$ as in equation~(\ref{beta}), we 
numerically solve the steady-state transport equation. 
 { We set $p_0$ to be $32$ keV/c, which corresponds to an injection energy of $1$ keV. } 
We use the same shock profile as \citet{Drury.etal82}, given by equation~(\ref{eq:u})
and assume a compression ratio of $3.5$ (thus a strong shock).
Note that the outflow plasma speed at a reconnection site
is $\sim V_A$. Therefore for a shock with a compression ratio
of $3.5$,  $u_1-u_2$ in equation~(\ref{eq:kappa0}) is $\sim 10^4$ km/s.
We use $\beta_0 = 0.2$. 
\citet{Tsuneta.Naito98} have used the Bohm approximation for $\kappa$.
With the Bohm approximation $\kappa = \frac{1}{3} v R_l$ and the above values for a typical flare site,
then a $300$ keV  electron will have $\beta_s = 0.2 $  and a $2$ MeV electron will have
$\beta_s = 1.0$, suggesting our choice of $\beta_0 = 0.2$ is reasonable.

\begin{figure}[ht]
\includegraphics[width=0.9\textwidth]{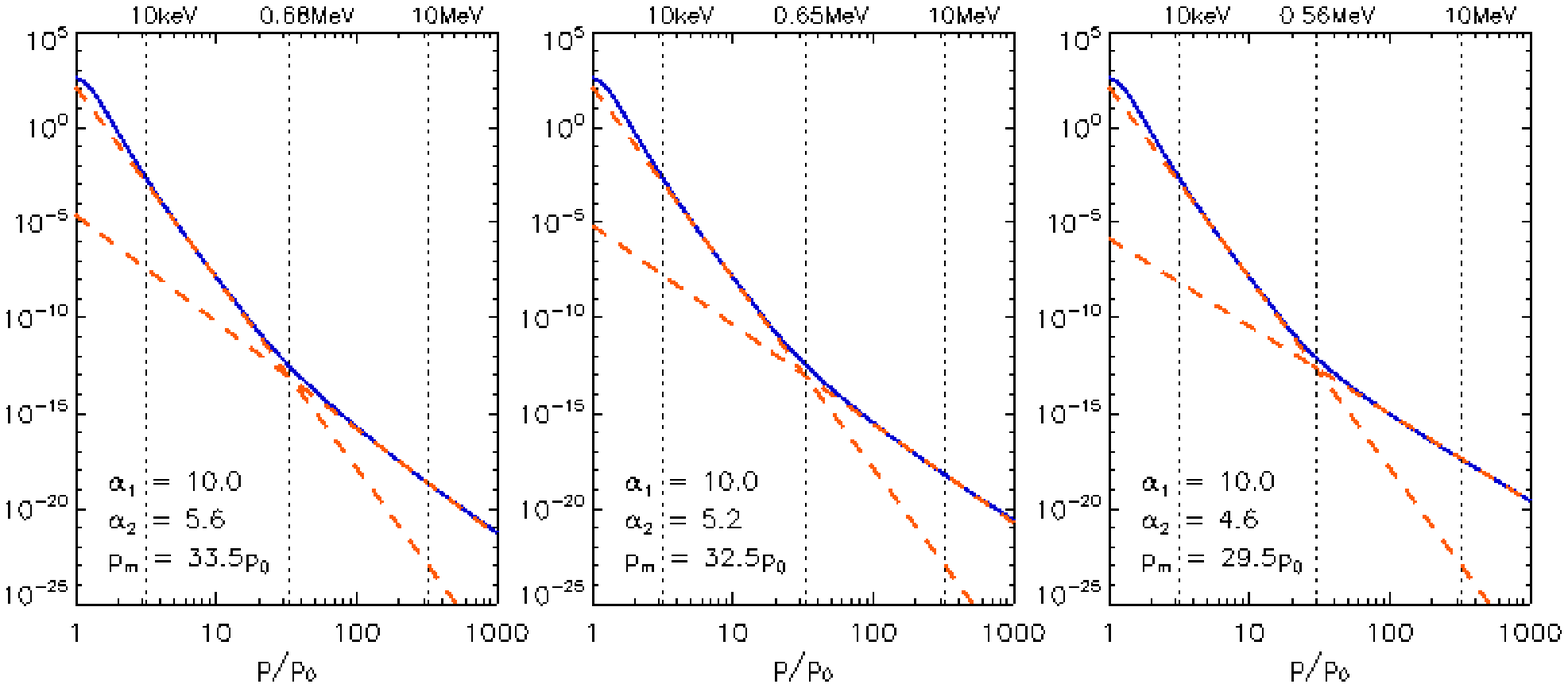}
\caption{The electron spectrum for three cases: i: Kolmogorov-like inertial range;
ii: IK-like inertial range; iii: Bohm diffusion Approximation. }
\label{fig:spectra}
\end{figure}

Figure~\ref{fig:spectra} plots the steady state  electron spectrum
for three cases that have different inertial range turbulence
spectrum: i): Kolmogorov-like; ii): IK-like; and iii): Bohm diffusion approximation. 
In each panel, the two dashed lines are power law fittings  { $\sim
(p/p_{0})^{-\alpha}$}, with $\alpha_1$ and $\alpha_2$  the fitted
spectral indices to the spectrum at the low and high energies
respectively, and $p_m$ the fitted break momentum.
We set $p_b$ to be  $13 p_0 \sim 0.416$ MeV/c in the simulation, 
which corresponds to a $E_b$ of $0.15$ MeV. 
Note that $p_m$ is larger than $ p_b = \gamma m_e \Omega/k_b $ by about a factor of $\sim 2$.
Figure~\ref{fig:spectra} is the most important result of this paper. 
It shows that diffusive shock acceleration at a
finite-width termination shock in solar flares can naturally lead
to a hardening of the accelerated electron spectrum.

\section{Discussions and conclusions}
Clearly, the hardening requires the following conditions to be met. 
First, the existence of a termination shock at flare site with a finite shock width 
$L_{diff} \sim c/{\omega_{pi}}$.
Second, the diffusion coefficient $\kappa$ needs to be close to a constant at
low energies and increases with electron energy at high energies.
Third, it is necessary that $\kappa < \Delta U L_{diff} $ at
energies below the break and $\kappa > \Delta U L_{diff}$ at
energies above the break. 

For any given flare, none of these conditions are necessarily satisfied. 
 
\red{Consider the first condition. While it is hard to identify a termination shock at a flare observationally, there are 
indirect clues of such shocks. For example, type II radio bursts without frequency drift has been used 
by \citet{Warmuth.etal09} to infer the existence of flare termination shocks. 
Further observational evidence of flare termination shock, and in particular its size, are welcomed.}

\red{For the second condition: if electrons resonate with the dissipation range of the turbulence 
through cyclotron resonance and that the dissipation range has a power spectrum  $I(k) \sim k^{-2.7} $, 
then  we find $\kappa$ is indeed close to a constant at low energies and 
increases with electron energy at high energies.
While in-situ solar wind observations do suggest such a  $\sim k^{-2.7} $ dissipation range,
direct confirmation of such a $k$ dependence in the flare site is impossible.}

To satisfy the third condition will put a strong constraint on the turbulence level at the flare, which 
can vary much from one to another. Consequently, the hardening does not occur in all flares. 
For example, if for a given flare,  $\beta_s \sim 1$ instead of $\beta_s \sim 0.2$ at lower energies, 
then there will be no hardenings, even if both conditions 1 and 2 are satisfied. 
Because a larger $\beta_s$ implies a larger $\kappa$, therefore a less efficient acceleration,
so one implication of our proposed scenario is the following: Events where the continuum emission extend to very high
energies (efficient acceleration) likely show hardenings and have
softer spectra at lower energies, and events where the continuum
emissions do not extend to high energies (inefficient acceleration) likely have harder spectra 
at low energies than those events that extend to higher energies.

Another consequence of our proposal is the correlation between the low energy photon spectral index $\gamma_1$ and the 
break momentum $p_m$. Consider two nearly identical flares A and B except that flare A
has a larger $k_b$ (i.e. the inertial range in flare A extends to a smaller scale). 
Then $\beta_s$ at $p < p_m$ for flare A is smaller. Therefore,
$p_m$ and $\alpha_1$ are anti-correlated. Observations do show such a anti-correlation and this 
is discussed in details in Kong et al., (in preparation).

In summary, we offer an explanation for the observed continuum spectral hardening in solar flares that is
based on DSA. To our knowledge, no previous works have addressed the hardening of emission spectrum explicitly.
Further observational and theoretical studies along the proposed mechanism will be pursued in future works.

\acknowledgments This work is supported in part by NSF grants
ATM-0847719, AGS1135432, and NASA grants NNH07ZDA001N-HGI and NNX11AO64G at UAHuntsville
and the 973 Program No. 2012CB825601 and NNSFC Grant Nos. 41274175 and 41028004 at SDUWH.
XLK acknowledges financial support by the Shandong University Graduate Study Abroad Fund.

\end{document}